\documentclass[12pt,a4paper]{article}
\usepackage{epsfig}
\usepackage{pml}
\usepackage{times}

\newcommand{\captionfonts}{\small}

\makeatletter  
\long\def\@makecaption#1#2{%
  \vskip\abovecaptionskip
  \sbox\@tempboxa{{\captionfonts #1: #2}}%
  \ifdim \wd\@tempboxa >\hsize
    {\captionfonts #1: #2\par}
  \else
    \hbox to\hsize{\hfil\box\@tempboxa\hfil}%
  \fi
  \vskip\belowcaptionskip}
\makeatother   

\begin{document}
\begin{center}\large\bf
  Dislocation nucleation and vacancy formation during high-speed
  deformation of fcc metals
\end{center}
\begin{center}\sc
  J. Schi{\o}tz,$^1$ T. Leffers$^2$ {\rm and} B. N. Singh$^2$
\end{center}
\begin{center}
  $^1$ Center for Atomic-scale Materials Physics (CAMP) and Department 
  of Physics, Technical University of Denmark, DK-2800 Kongens Lyngby, 
  Denmark\\
  $^2$ Materials Research Department, Ris{\o} National Laboratory,
  DK-4000 Roskilde, Denmark
\end{center}
\begin{center}\footnotesize
\end{center}
\begin{abstract}
  Recently, a dislocation free deformation mechanism was proposed by
  Kiritani \emph{et al.}, based on a series of experiments where thin
  foils of fcc metals were deformed at very high strain rates.  In the
  experimental study, they observed a large density of stacking fault
  tetrahedra, but very low dislocation densities in the foils after
  deformation.  This was interpreted as evidence for a new
  \emph{dislocation-free} deformation mechanism, resulting in a very
  high vacancy production rate.
  
  In this paper we investigate this proposition using large-scale
  computer simulations of bulk and thin films of copper.  To favour
  such a dislocation-free deformation mechanism, we have made
  dislocation nucleation very difficult by not introducing any
  potential dislocation sources in the initial configuration.
  Nevertheless, we observe the nucleation of dislocation loops, and
  the deformation is carried by dislocations.  The dislocations are
  nucleated as single Shockley partials.

  The large stresses required before dislocations are nucleated result 
  in a very high dislocation density, and therefore in many inelastic
  interactions between the dislocations.  These interactions create
  vacancies, and a very large vacancy concentration is quickly reached.
\end{abstract}

\section{Introduction}
\label{sec:intro}

In a recent paper, Kiritani \emph{et al.} reported that a large number
of vacancies were produced during high-speed heavy plastic deformation
of thin foils of fcc metals \cite{KiSaKiArOgArSh99}.  They observed a
large density of stacking-fault tetrahedra but very low dislocation
densities in the foils after deformation.  As a possible explanation,
they propose a dislocation-free deformation mechanism.

In this paper we use computer simulations to investigate, whether such a
dislocation-free deformation mechanism becomes possible at very high
strain rates, and we look for other possible explanations of the high
rate of vacancy production.  We have attempted to make the generation
of dislocations as difficult as possible in order to favour such a
dislocation-free deformation mechanism.  We have therefore chosen to
simulate systems that are initially dislocation free (i.e. without
conventional dislocation sources such as Frank-Read sources).  We have
also chosen to simulate single crystals in order to avoid dislocation
emission from grain boundaries.  Simulations of high-speed deformation 
of polycrystalline (nanocrystalline) material have been reported
elsewhere \cite{ScDiJa98,ScVeDiJa99,SwCa98}.

\section{Simulation methods and setup}
\label{sec:simul}

The simulations were performed using molecular dynamics.  The
interatomic interactions were described using the effective medium
theory \cite{JaNoPu87,JaStNo96} which gives a good description of the
metals investigated here.  The crystal is deformed at a constant
strain rate, while keeping the stress in the transverse direction
approximately constant, as described in details by
\citet{ScVeDiJa99}.  In all cases a strain rate of $10^9 s^{-1}$ was
used (versus approximately $10^5 s^{-1}$ in the experiments).  To keep
the temperature of the system approximately constant during the
simulation, Langevin dynamics are used, i.e.\ a friction and a
fluctuating force are added to the equations of motion of the atoms
\citep[see e.g.][]{AlTi87}.  We use a timestep of 5\,fs, safely below
the value where the dynamics become unstable.

During the simulations, configurations were rapidly quenched and
common neighbour analysis \cite{HoAn87,FaJo94} was used to identify the 
local crystal structure.  This was used to generate plots where all atoms
except atoms at the dislocation cores were made invisible, allowing
visualization of the dislocation structures.

Four different setups were used for the simulations.

\subsection{System I.}
\label{sec:sysI}

The first system investigated was an fcc single crystal containing
approximately 95\,000 atoms.  The system is approximately cubic with a
side length of 105\,\AA.  There are periodic boundary conditions along
all three directions to emulate that the system is far from any
surfaces.  The crystal structure is perfect, apart from four randomly
introduced vacancies.  They were introduced in case the vacancies
might participate in the proposed dislocation free deformation
mechanism.  The crystal is deformed along the [521] direction.  This
direction was chosen to avoid high-symmetry directions.

\subsection{System II.}
\label{sec:sysII}

Similar to system I, but with free boundary conditions in the
directions perpendicular to the pulling direction, i.e.\ the geometry
is that of an infinite wire with square cross section.

\subsection{System III.}
\label{sec:sysIII}

A single fcc crystal with approximately 765\,000 atoms.  The system is
approximately cubic with a side length of 210\,\AA.  As in system I
there are periodic boundary conditions in all three directions. The
crystal is deformed along the [$\bar{\mbox{1}}$05] direction.

\subsection{System IV.}
\label{sec:sysIV}

As system III, but the periodic boundaries in one direction were
replaced with free boundary conditions.  This results in a ``film''
geometry, a thin film (thickness 210\,\AA) with free (512) surfaces,
repeated infinitely in the two remaining directions.  The film was
deformed along the [$\bar{\mbox{1}}$05] direction.

As the preexisting vacancies in systems I and II did not participate
in the deformation, no vacancies were introduced in the initial
configuration of systems III and IV.

\section{Results}
\label{sec:results}

\begin{figure}[tp]
  \begin{center}
    \epsfig{file=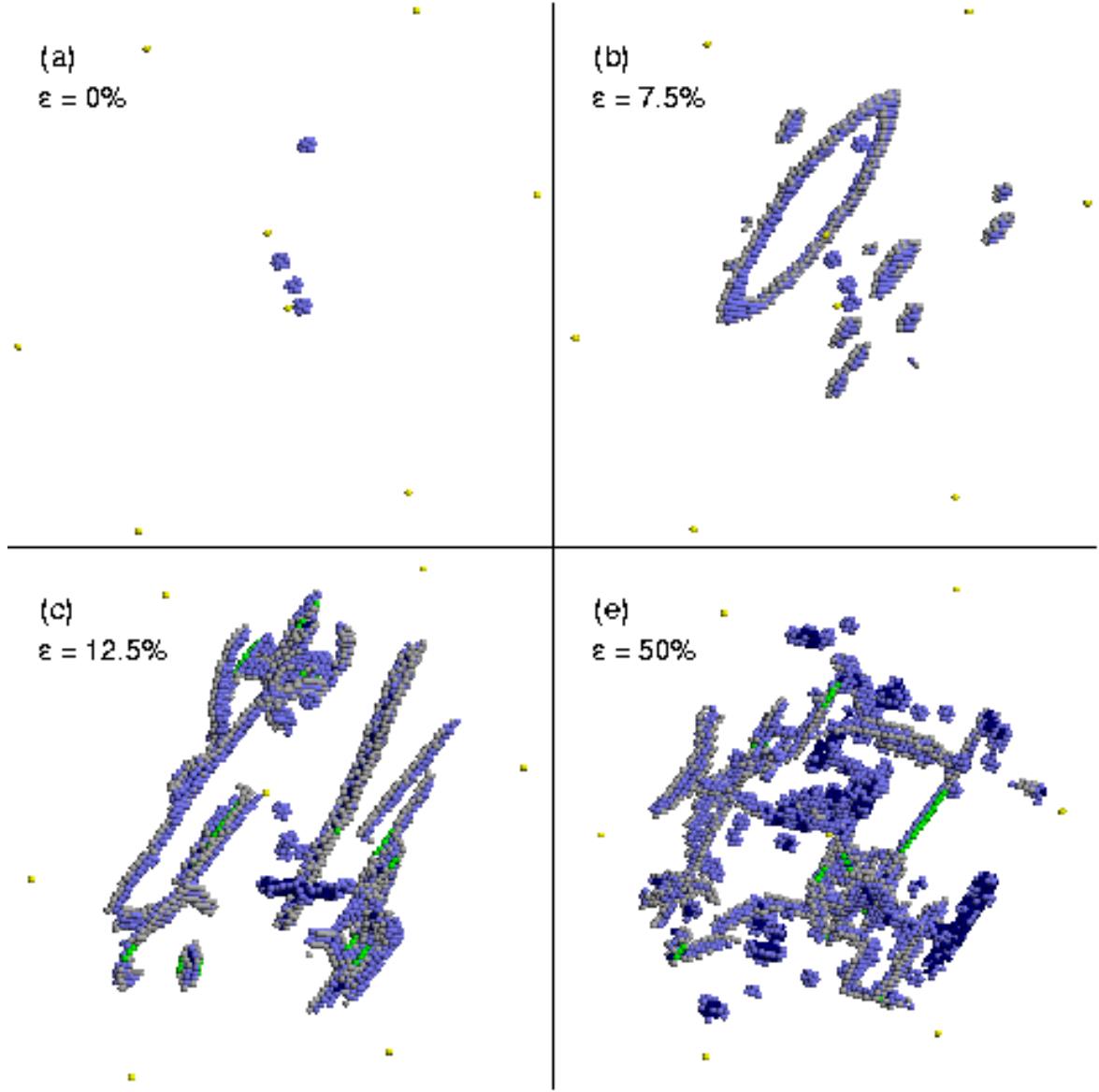, width=\linewidth}
    \caption{System I.  In this and the following figures only atoms
      near crystal defects are shown.  White atoms have a coordination
      number of 12, blue atoms have a coordination number of 11, dark
      blue atoms have coordination numbers below 11, and green atoms
      have a coordination number of 13.  Eight yellow atoms are used
      to mark the corners of the simulation cell.  In \emph{part (a)}
      the four initial vacancies are seen.  The vacancies themselves
      are not shown, but the 12 atoms next to the vacancies are shown
      in blue as they are missing a nearest neighbour and therefore
      have the coordination number 11.  In \emph{part (b)} the first
      dislocation activity is seen.  \emph{Part (c)} shows the first
      generation of vacancies (the string of atoms with coordination
      number 11 or below).  \emph{Part (d)} shows the final
      configuration, where a high density of vacancies is seen.}
    \label{fig:sysI}
  \end{center}
\end{figure}

Figure \ref{fig:sysI} illustrates the deformation of System I.  The same
deformation mode is seen in system III and (with a small modification, 
see below) in system IV.

In figure \ref{fig:sysI}(b) the first dislocation activity is observed
in the form of a few homogeneously nucleated dislocation loops.  The
loops expand rapidly.  Due to the periodic boundary conditions the
dislocations cannot disappear from the sample.  They will continue to
move, causing significant plastic deformation.

After some dislocation activity has occurred, we observe the first
creation of vacancies in figure \ref{fig:sysI}(c).  A
``sausage-like'' string of vacancies have appeared in the middle of
the system.  As the deformation proceeds, the string of vacancies is
broken up by the dislocation activity, and the vacancies are dispersed 
into single vacancies and smaller clusters.  At the same time, more
vacancies are being generated.

\subsection{Dislocation nucleation.}

\begin{figure}[tbp]
  \begin{center}
    \epsfig{file=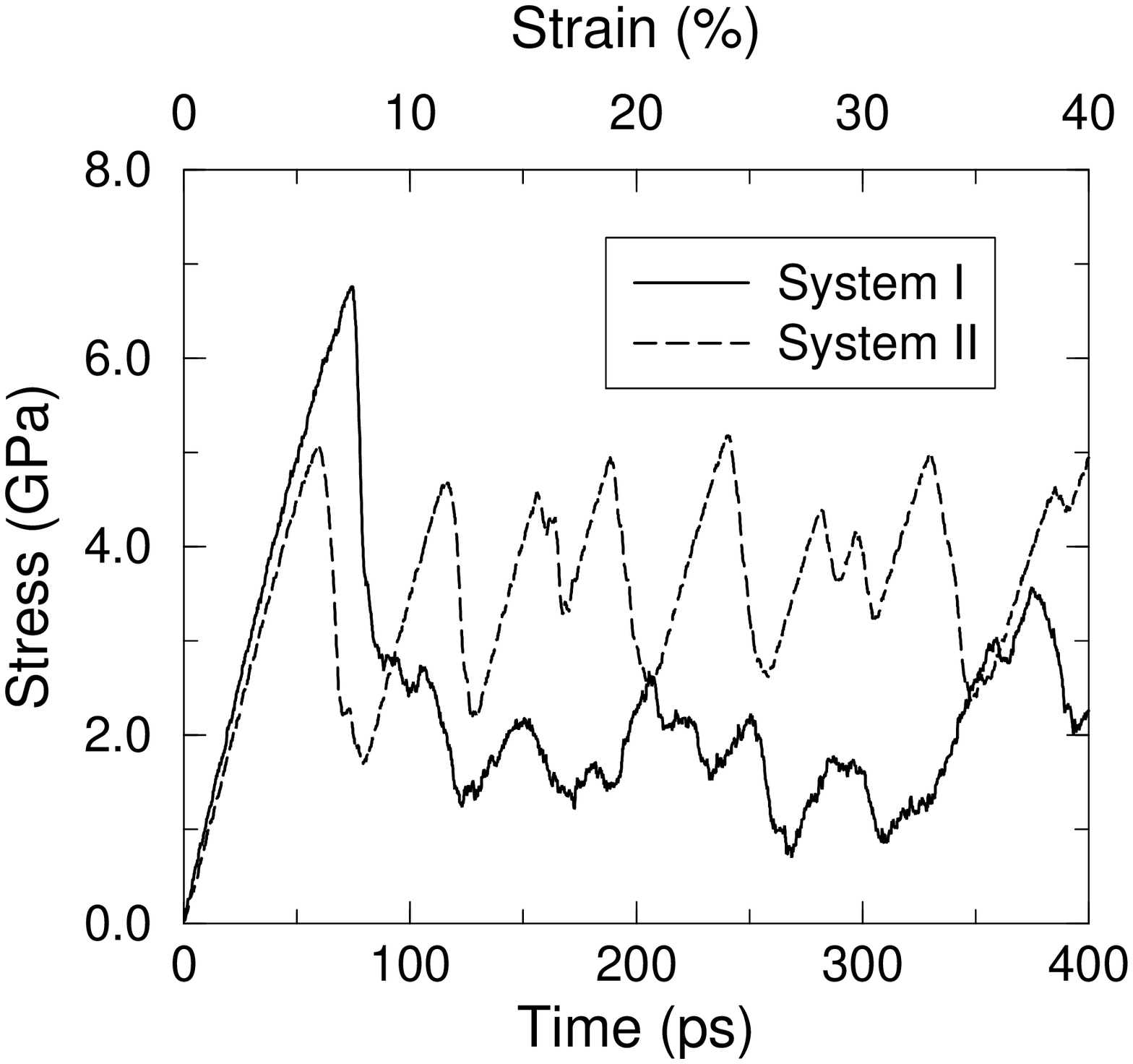, width=0.48\linewidth}\hfill
    \epsfig{file=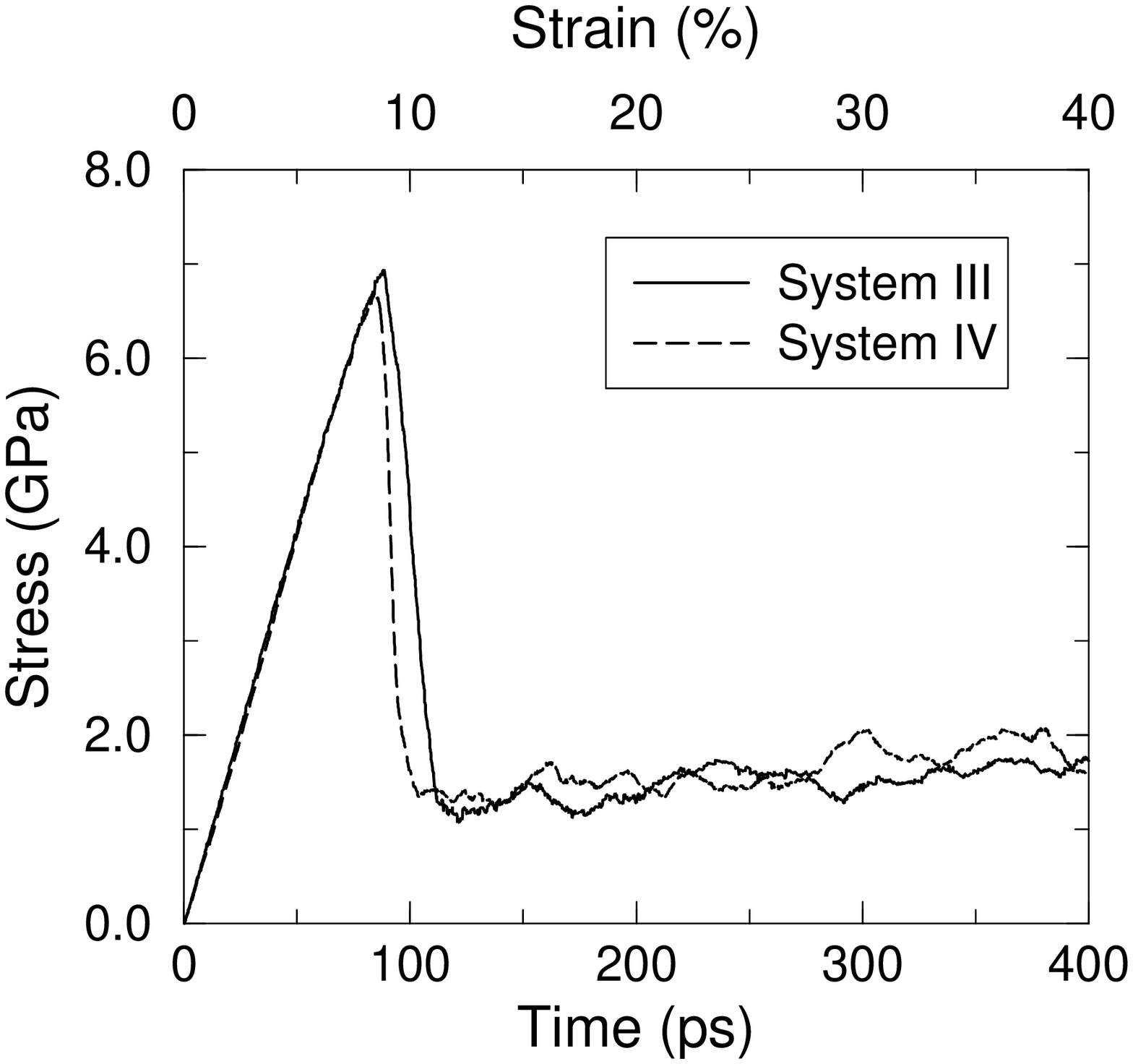, width=0.48\linewidth}
    \caption{Stress versus strain and time for the four simulations
      mentioned in the text.  Systems I, III and IV behave in very
      similar ways, once the stress reaches a critical level
      dislocations are nucleated.  As the dislocations cannot escape
      the system due to the periodic boundary conditions, a much lower 
      flow stress is rapidly reached.  In system II the dislocations
      can leave the system at the surfaces, and the stress has to
      build up again to nucleate new dislocations.  The fluctuation
      during the flow phase is significantly larger in system I
      compared to system III and IV, this is because systems III and
      IV contain eight times as many atoms, so a better spatial
      averaging of the stress is obtained.}
    \label{fig:stress}
  \end{center}
\end{figure}

Figure \ref{fig:stress} shows the stress versus time and strain for
the simulations.  It is clearly seen that the deformation falls in two 
parts, an elastic and a plastic regime.  In the elastic regime (below
$\varepsilon \approx 7.5\%$), no dislocations are present in the sample, 
and no plastic deformation is possible.  As there are no dislocation
sources, dislocations cannot be nucleated until the stress is close to 
the theoretical shear stress, where (111) planes can glide with
respect to each other.  At this point thermal fluctuations are enough
to nucleate dislocations.  This is seen in figure \ref{fig:sysI}b.  A
number of loops have been nucleated almost simultaneously.  The first
one to be nucleated is clearly seen, but a number of emerging loops are 
seen as well.

The loops consist of a single Shockley partial dislocations, and are
thus faulted loops.  The second partial is not being nucleated.  This
is because the barrier for nucleating the second partial is almost as
high as the barrier for nucleating the first partial, but by the time
the first partial is nucleated, it screens the stress field.  The
trailing partials are thus not nucleated.  It should also be noted
that the stacking fault energy in copper is relatively low, and
furthermore the stacking fault energy in the simulation is
approximately a factor of two below the experimental value
\cite{ScCa00}.  For similar reasons, only Shockley partial
dislocations are seen in simulations of the mechanical deformation of
nanocrystalline metals \cite{ScVeDiJa99}.

\begin{figure}[tbp]
  \begin{center}
    \epsfig{file=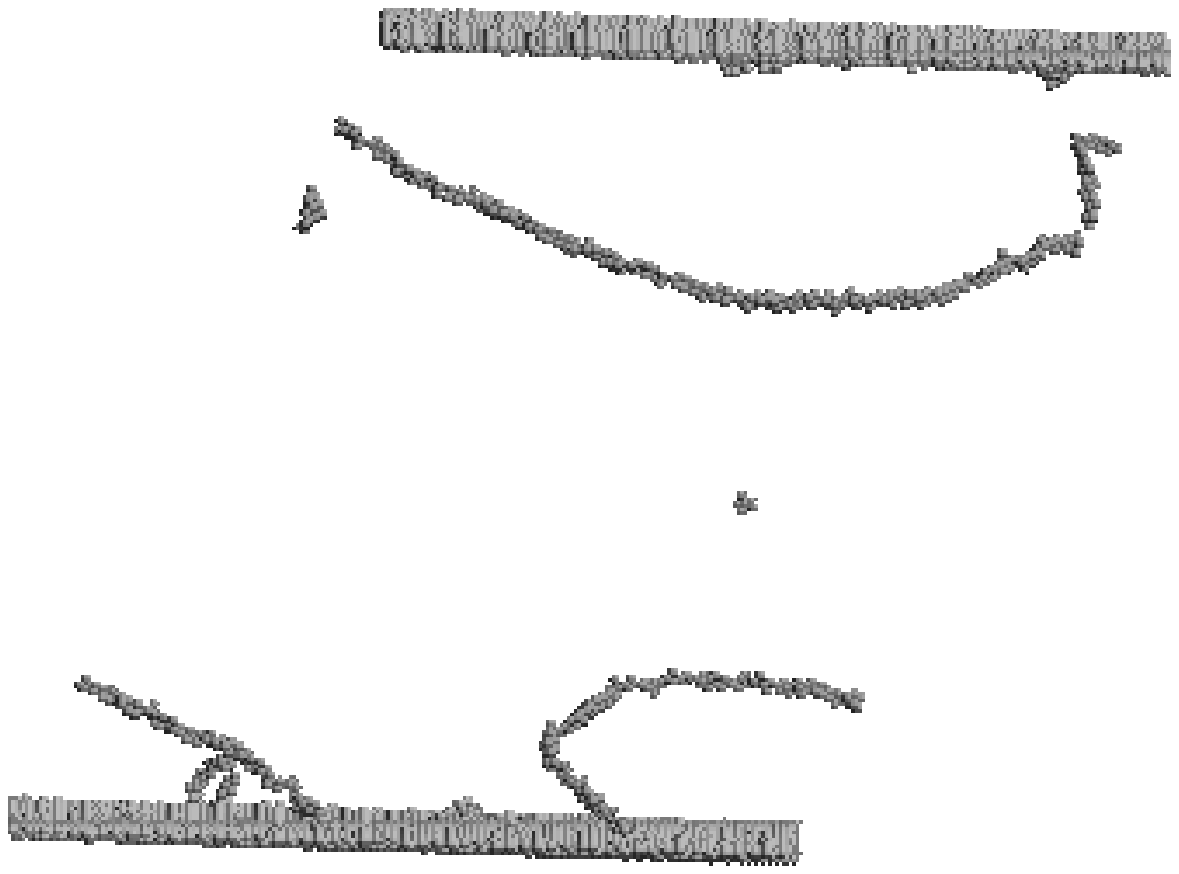, height=0.29\linewidth, clip=}\hfill
    \epsfig{file=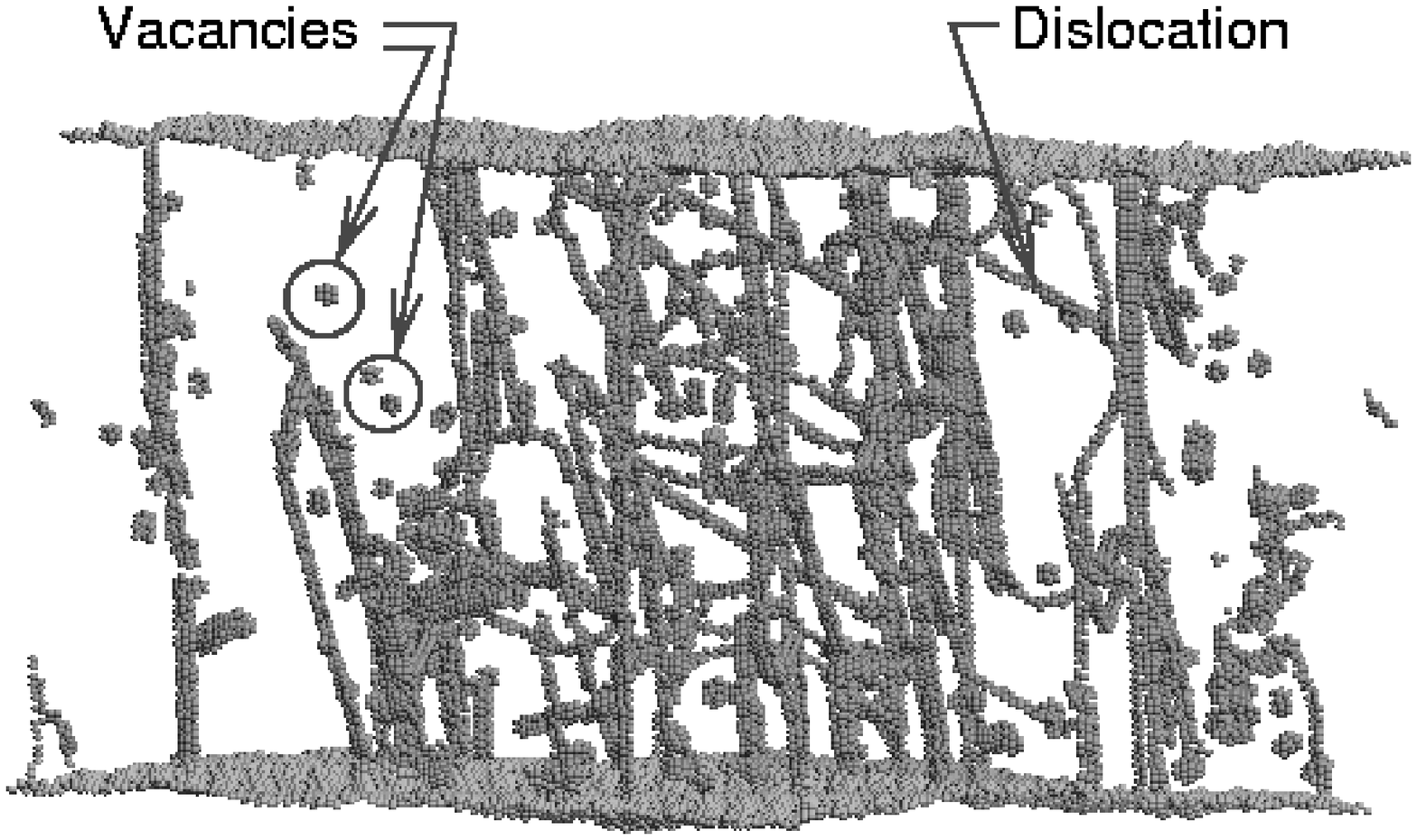, height=0.33\linewidth, clip=}
    \caption{System IV.  \emph{Left:} just after the nucleation of the 
      first dislocations.  The dislocations are seen to be nucleated
      at the surfaces.  \emph{Right:} After 40\% deformation, large
      densities of dislocations and vacancies are seen.  Note that to
      facilitate the visualization, the system is shown from different 
      angles in the two figures.}
    \label{fig:sysIV}
  \end{center}
\end{figure}

The behavior in system I and III are very similar, as they only differ 
by the size of the system.  In system IV, where free surfaces are
present, the dislocations are (perhaps not surprisingly) nucleated at
the surfaces.  Since the surfaces are defect-free, they are not good
sources for dislocations, and the dislocations are nucleated only
marginally earlier than in the bulk simulations (see
fig.~\ref{fig:stress}).  The dislocations are again nucleated as
single Shockley partials in the form of a faulted half-loop.  When the 
loop reaches the opposite surface the result is two dislocations that
extend from one surface to the other.   Such dislocations cannot run
out of the sample due to the periodic boundary conditions, and again a 
large dislocation density is the result.  See fig.~\ref{fig:sysIV}.

In system II dislocations are nucleated at the corners of the system.
Again, the lack of defects prevents dislocation nucleation until a
very high stress is reached, even though the lower atomic coordination at
the corners allows dislocation nucleation at a stress that is
approximately 25\% lower than in the bulk and film configurations.
Unlike the other geometries, there is nothing to prevent the
dislocations from moving out of the sample at the opposite surface,
and this is indeed what happens.  Initially, two dislocations are
nucleated approximately simultaneously on opposite corner.  Just
before the dislocations reach the opposite corner new dislocations are
nucleated, perhaps due to the local heating caused by the moving
dislocation.  This repeats itself a few times until the stress has
been reduced.  After the brief surge of dislocation activity, the
sample is again dislocation free and the stress can build up again
(fig.~\ref{fig:stress}).  No evidence is seen for kinematic generation
of new dislocations near the rapidly moving dislocation.  This has
previously been seen in simulations under similar conditions except
the temperature was close to zero Kelvin \cite{ScJaNi95}.  The reason
appears to be that phonon drag at the temperature of the simulations
presented here (450K) prevents the dislocations from reaching the
Rayleigh velocity.

\subsection{Vacancy production.}

After some dislocation activity vacancies appear in the simulation.
The vacancies initially appear in the form of short strings of
vacancies.  A closer look at the simulation shows how the vacancy
structure appears.  Dislocation interactions have created dislocations
where segments are on different glide planes.  When these dislocations
meet, it is possible that two segments of edge dislocation with
opposite signs annihilate.  If the dislocations are not on the same
slip plane, but on adjacent slip planes, a string of vacancies or
interstitials is generated.  In the simulations interstitial
generation by this mechanism is never observed, probably because the
energy of a string of interstitials is so high that the colliding
dislocations pass by each other instead of annihilating.  The vacancy
formation mechanism is further discussed elsewhere \cite{ScLeSi00x}.

At the end of the simulation (figures \ref{fig:sysI}(d) and
\ref{fig:sysIV}) a large number of vacancies have been formed,
apparently by this mechanism.  The vacancy density at this point is
of the order of $2 \cdot 10^{-4}$.

\section{Discussion}
\label{sec:discussion}

The simulations presented here demonstrate that even in a 
situation where the generation of dislocations has been made as
difficult as possible, dislocation nucleation does occur before any
dislocation-free deformation mechanism is activated.  Once 
dislocations are nucleated, they carry the plastic deformation, and no 
evidence of a dislocation free deformation mechanism is seen neither
in bulk nor in thin film simulations.

The high vacancy concentration seen experimentally
\cite{KiSaKiArOgArSh99} is reproduced in the simulations.  The
vacancies are created by dislocation--dislocation interactions, which
raises the question ``why are a similar concentration of vacancies not 
seen in ordinary deformation experiments?''  After all, the same
number of dislocations must pass through the sample to give the same
final strain, regardless of the strain rate.   However, the number of
dislocation-dislocation interactions is proportional not only to the
number of dislocations passing through the sample, but also to the
average dislocation density.  The dislocation density is likely to be
significantly larger under high-speed deformation, partly because more 
dislocation sources may be activated by the higher stresses reached
during the high-speed deformation, partly because new dislocations are 
nucleated from the dislocation sources before the previous ones have
time to reach a dislocation sink.

The vacancy density observed in the simulations ($2 \cdot 10^{-4}$) is
in excellent agreement with the observed densities ($10^{-4}$).  To
some extent this agreement is fortuitous, since the strain rate is
much higher in the simulations than in the experiments, whereas the
total strain is higher in the experiments.  It is, however, important
that the right order of magnitude is found for the vacancy density.

In the experiments presented by \citet{KiSaKiArOgArSh99} no (or very
few) dislocations are seen in the sample after deformation --- in
constrast to the simulations, where a large dislocation density is
seen (figure \ref{fig:sysIV}).  However, the time scale of the
experiment is much longer than the time scale of the simulations,
giving ample time for the dislocations to disappear.  It is clear that
when the dislocation line goes from one surface to the other (as is
the case of most of the dislocations in figure \ref{fig:sysIV}) they
cannot simply move to the surface and disappear.  In the reported
experiments, the TEM micrographs were taken near the egde where the
foils were torn apart since this is where the foil was sufficiently
thin.  In this case it must be assumed that the foil is wedge-shaped,
and that the dislocations can leave the sample by moving to the edge
where the foil broke.  In \emph{in-situ} TEM studies, where the foil
is examined before it breaks \cite{Ki00x}, the thinnest part of the
film is examined.  If there are no dislocation sources in this part of
the film, the stress will increase until dslocations are nucleated at
the surface.  These dislocations will be accelerated to very high
velocities and leave the region being studied by the TEM.  They are
not observed because of their high velocity.  The phenomenon that
dislocations are rapidly expelled from a region with a local stress
concentration is well known from fracture mechanics, where a
dislocation free zone (DFZ) is formed in front of the crack tip
\cite{KoOh80,ChOh81,Oh85}.  We propose that a similar DFZ is formed
here, and that this explains why dislocations are not observed in the
experiments.

\citet{ShSuKi99} have recently published simulations of high-speed
deformation in Cu and Al (the interatomic potential used was not
reported).  They have used systems with 4000 atoms, and either full
periodic boundary conditions (similar to system I) or with
wire-symmetry (as system II).  They report that the deformation
occurs without dislocations, but by ``tilting of atom rows''.
However, this tilting is nucleated at one surface and proceeds through 
the specimen.  Based on the published material, we are not able to see 
the difference between this mechanism and the propagation of a
dislocation with screw character through the specimen.

In simulations of nanocrystalline metals a ``dislocation-free''
deformation mechanism is seen where the grain boundaries carry the
deformation \cite{ScVeDiJa99,ScDiJa98}.  Such a mechanism can be ruled
out in the experiments by \citet{KiSaKiArOgArSh99} as it only works
when the density of grain boundaries is extremely high, i.e.\ at grain
sizes below approximately 10 nm.  Furthermore, no generation of
vacancies is expected.

\section{Conclusions}

We have simulated high-speed deformation of copper in bulk and in thin
films.  In spite of not having included any dislocation sources in the
initial configuration, dislocations are nucleated and the plastic
deformation occur by the motion of dislocations.  We thus find no
evidence for the dislocation-free deformation mechanism proposed by
\citet{KiSaKiArOgArSh99}.  Inelastic interactions between the
dislocations results in the production of vacancies.  After the
deformation is complete, a very high vacancy density is seen, in
agreement with the experimental observations.  We propose that the
reason why no dislocations are seen in the experiments is the
formation of a ``dislocation free zone'' where the dislocations
rapidly leave the part of the film where the stresses are largest.

\section*{Acknowledgments}

We would like to thank Prof.~Kiritani for fruitful discussions.  The
Center for Atomic-scale Materials Science (CAMP) is sponsored by the
Danish National Research Foundation.  This work was done as a
collaboration between CAMP and the Engineering Science Center for
Structural Characterization and Modelling of Materials at Materials
Research Department, Ris{\o}.  Parallel computer time was financed by
the Danish Research Councils through grant no.~9501775.  The authors
are collaborating with the Academic Frontier Research Center for
Ultra-high Speed Plastic Deformation at Hiroshima Institute of
Technology.

\section*{References}


\end{document}